\documentclass[10pt,conference]{IEEEtran}
\usepackage{amsmath,amssymb,amsthm,bm}
\usepackage{cite}
\usepackage{graphicx}
\usepackage{booktabs}
\usepackage{amsmath, amssymb, amsthm}
\usepackage{algorithm}
\usepackage{algorithmic}
\theoremstyle{definition} 
\newtheorem{definition}{Definition}
\newtheorem{theorem}{Theorem}
\newtheorem{proposition}{Proposition}
\usepackage{fancyhdr}
\fancypagestyle{firstpage}{%
  \fancyhf{}%
  \fancyfoot[C]{\parbox{\textwidth}{\scriptsize\setlength{\baselineskip}{8pt}%
  \copyright\ 2026 IEEE. Personal use of this material is permitted. Permission from IEEE must be obtained for all other uses, including reprinting/republishing this material for advertising or promotional purposes, collecting new collective works for resale or redistribution to servers or lists, or reuse of any copyrighted component of this work in other works.}}%
}

\begin{document}

\title{Latency-Aware Uplink NOMA-MEC with PASS:
A Feasibility-First Resource Allocation Framework}

\author{
\IEEEauthorblockN{Sepehr Abolhasani \IEEEauthorrefmark{1}, Saba Asaad \IEEEauthorrefmark{2}, S. Mohammad Razavizadeh \IEEEauthorrefmark{1}, and Jiguang He \IEEEauthorrefmark{3}}
\IEEEauthorblockA{
\IEEEauthorrefmark{1} \textit{School of Electrical Engineering,
Iran University of Science and Technology (IUST),
Tehran, Iran} \\
\IEEEauthorrefmark{2} \textit{Department of Electrical and Computer Engineering, University of Toronto, Toronto, Canada} \\
\IEEEauthorrefmark{3} \textit{School of Computing and Information Technology, Great Bay University, Dongguan 523000, China}
}}
\maketitle

\thispagestyle{firstpage}

\begin{abstract}
Meeting hard task-completion deadlines for many simultaneously offloading users is
the binding challenge in mobile-edge computing (MEC). As the user count grows, the
transmit power required to satisfy every deadline within per-user budgets rises
until no feasible allocation exists. We take a \emph{feasibility-first} view of this $K$-user problem in an uplink non-orthogonal multiple access (NOMA) MEC
system served by a pinching-antenna system (PASS), whose radiating points establish
strong line-of-sight (LoS) links that lower the received-power requirement. We
formulate a latency-constrained joint optimization of transmit powers, CPU
frequencies, and antenna configuration. We then derive a \emph{closed-form
least-power feasibility test}: the componentwise-least allocation meeting every
deadline. It certifies feasibility with the largest attainable margin to the power
budgets and defines a practical \emph{quality-of-service (QoS)-feasibility outage} metric, the
probability that no allocation meets all deadlines within budget. We solve the problem with a three-block alternating-optimization algorithm that is provably monotone-convergent and has polynomial complexity. Numerical results show that the proposed
scheme sustains deadline feasibility at user loads where equal-power,
orthogonal multiple access (OMA), and single-antenna baselines fail or degrade sharply, each by a mechanism predicted directly from the system model, whereas the proposed design degrades gracefully.
Stress tests further show that
feasibility is sustained under order-of-magnitude budget reductions, and that the
pinching-antenna array itself enables constrained-budget operation. The result is a
scalable, closed-form-initialized design for deadline-feasible offloading.
\end{abstract}


\section{Introduction}\label{sec:intro}
The proliferation of Internet-of-Things (IoT) devices and latency-sensitive applications
has imposed demands on communication and computation resources.
Mobile-edge computing (MEC) addresses the limited computational capability of such
devices by offloading tasks to edge servers co-located with base stations (BSs),
thereby reducing processing latency and backhaul overhead~\cite{wang2022irs_mec_noma}.
Under hard per-task deadlines, however, the coexistence of many devices sharing
limited spectrum makes it increasingly difficult to complete every task in
time: as the user count grows, the transmit power required to meet all deadlines within per-user budgets increases to the point where no feasible allocation exists. Sustaining this deadline feasibility as the system scales, rather than optimizing
efficiency alone, is the central challenge we address.

Non-orthogonal multiple access (NOMA) mitigates the spectrum bottleneck by letting
users share the same time-frequency resource via superposition coding and successive
interference cancellation (SIC), thereby improving aggregate offloading
throughput~\cite{wang2022irs_mec_noma}. Its performance is
nonetheless limited by the user--BS channel quality, which can be severely degraded by path loss and blockage, especially at high frequencies. Reconfigurable intelligent
surfaces (RIS) reshape the propagation environment passively but may suffer from a double
path-loss effect and often require a large number of elements to provide substantial gain~\cite{wang2022irs_mec_noma}. Fluid and movable
antennas reposition elements within a limited region~\cite{zuo2024fluid_mec,zhu2024movable}, but their wavelength-scale range limits their ability
to combat large-scale path loss or serve widely distributed users.

Pinching-antenna systems (PASS) have recently emerged as a versatile evolution of
flexible-antenna technology~\cite{fukuda2022pinching}. Dielectric
waveguides are deployed across the coverage area, and small dielectric clips
(pinching antennas) radiate at configurable positions along them. Because the
waveguide loss is far below free-space path loss at the same frequency, a
strong line-of-sight (LoS) link can be established to nearby
users by positioning the radiating points~\cite{liu2026pass_tutorial}, making PASS inherently robust
against blockage and path loss~\cite{tyrovolas2025performance}. Recent works have studied
uplink rate maximization for PASS~\cite{tegos2025uplink_pass} and energy-efficiency maximization in
NOMA uplink PASS~\cite{zeng2025ee_noma_pass}; learning-based energy minimization and
delay minimization for pinching-antenna NOMA-MEC~\cite{ma2026drl_pass_mec,ai2026delay_pass_mec}
and deep reinforcement learning (DRL)-based latency minimization for PASS-assisted MEC~\cite{hu2026pass_mec} have
also been reported.

Closest to our setting, two recent works study pinching-antenna NOMA-MEC:
\cite{ma2026drl_pass_mec} solved a two-user, single-antenna case using DRL to minimize energy, and~\cite{ai2026delay_pass_mec}
minimized the maximum task delay via bisection-based alternating optimization. To
our knowledge, a general framework for the $K$-user case remains unexplored, and these works
optimize energy or delay rather than characterizing \emph{feasibility}, the ability to meet every deadline within budget, as the user count scales. Neither
provides a closed-form least-power feasibility characterization or an outage characterization, and both rely
on numerical or learned solvers rather than an analytical guarantee. This
combination is compelling precisely for feasibility: PASS lowers each user's
received-power requirement, NOMA enables simultaneous access, and provisioning radiating points and configuring powers and computation can extend the range of user counts for which every
deadline can be met. Our contributions are as follows.
\begin{itemize}
  \item We formulate a latency-constrained joint optimization of transmit powers,
  CPU frequencies, and pinching-antenna configuration for uplink PASS-NOMA-MEC under
  hard task-completion deadlines, with energy efficiency as the objective and the
  deadlines as feasibility constraints.
  \item We derive a closed-form \emph{least-power feasible initialization}
  (Proposition~\ref{prop:init}): the componentwise-least allocation meeting every
  deadline for a given system configuration, which certifies feasibility with the
    largest margin to the power budgets and yields an exact pure-offload feasibility test
  for that configuration. This test provides the basis for the QoS-feasibility
  outage metric used throughout.
  \item We solve the problem by a three-block alternating-optimization algorithm that
  guarantees monotonic improvement of the objective and has polynomial complexity, and whose least-power initializer provides an effective starting point
  in the deadline-binding regime.
  \item We show by simulation that the proposed design sustains zero observed
  QoS-feasibility outage over a wide range of user counts and degrades gracefully,
  tolerates order-of-magnitude budget reductions, and that the antenna array, not the
  power allocation alone, enables constrained-budget operation. Across all three
  studies it remains feasible in regimes where equal-power, OMA, and single-antenna
  baselines have already collapsed.
\end{itemize}
The remainder of this paper is organized as follows. Section~\ref{sec:model}
presents the system model and problem formulation; Section~\ref{sec:alg} develops the algorithm;
Section~\ref{sec:results} reports simulation results; Section~\ref{sec:concl}
concludes.

\section{System Model and Problem Formulation}\label{sec:model}
\begin{figure}[t]
\centering
\includegraphics[width=\columnwidth]{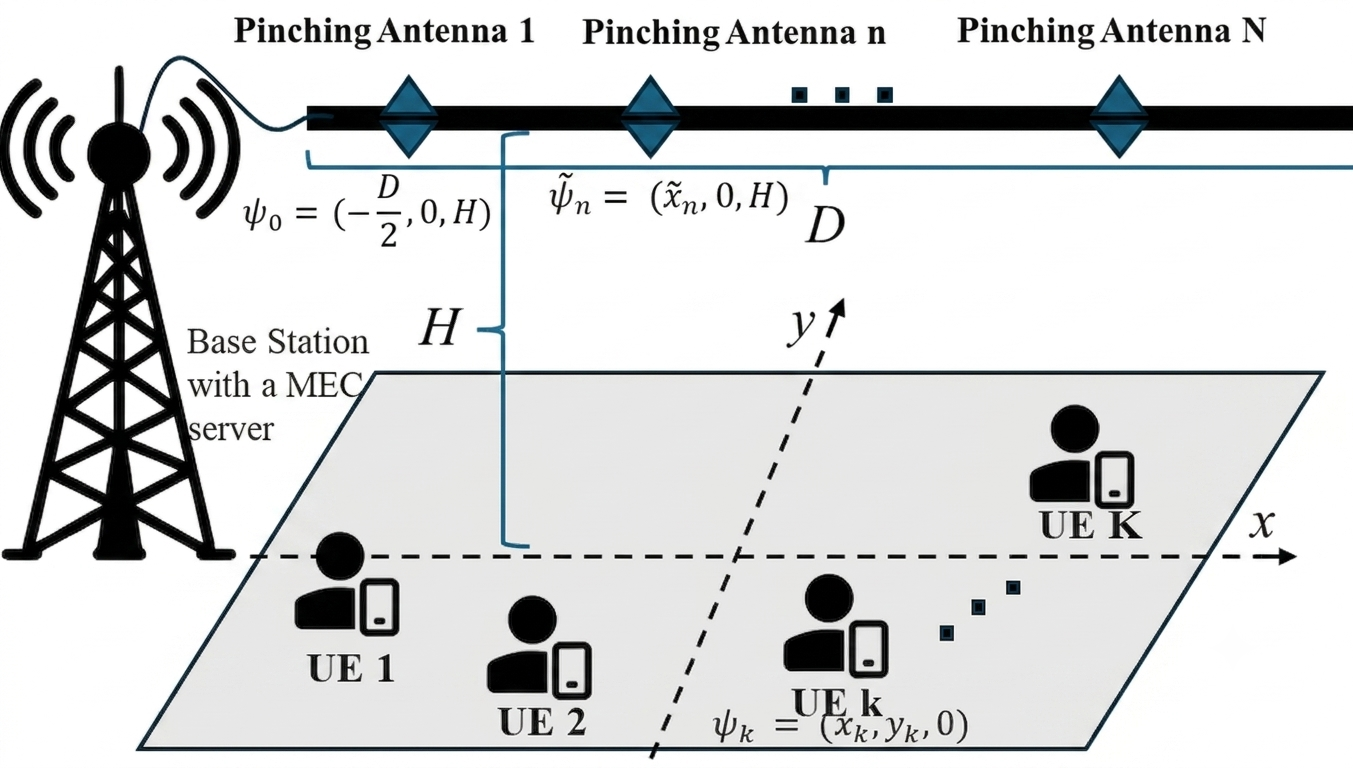}
\caption{Uplink PASS-NOMA-MEC system model.}
\label{fig:system}
\end{figure}

We consider an uplink system in which $K$ single-antenna users offload
latency-constrained computation tasks to a BS hosting an MEC server. The BS is
equipped with a PASS: a single dielectric waveguide of length $D$ along which $N$
pinching antennas (PAs) are deployed at configurable positions
$\tilde{\mathbf x}=[\tilde x_1,\ldots,\tilde x_N]^T$ and with configurable coupling
lengths $\boldsymbol\ell^{\mathrm c}=[\ell_1^{\mathrm c},\ldots,\ell_N^{\mathrm c}]^T$,
as illustrated in Fig.~\ref{fig:system}.
Because the PAs are physically actuated, the configuration is held fixed over a
reconfiguration epoch and optimized once per epoch.
\subsection{PA Channel Model}
Each PA extracts a fraction of the guided wave. The coupling length is restricted to $\ell_n^{\mathrm c}\in[0,\pi/(2\kappa)]$,
$\kappa$ being the waveguide coupling coefficient, so that the split coefficient
$\delta_n=\cos(\kappa\ell_n^{\mathrm c})$ sweeps monotonically over $[0,1]$: the
endpoint $\ell_n^{\mathrm c}=\pi/(2\kappa)$ gives $\delta_n=0$, the full-radiation
(matched) condition of~\cite{Asaad2026DynStatPASS}, while $\ell_n^{\mathrm c}=0$
gives $\delta_n=1$. On this branch, the map $\ell_n^{\mathrm c}\mapsto\delta_n$ is a bijection. The
effective split coefficient of the $n$-th PA~\cite{Asaad2026DynStatPASS} is
\begin{equation}\label{eq:An}
  A_n(\boldsymbol\delta)=\sqrt{1-\delta_n^2}\,\prod_{i=1}^{n-1}\delta_i,
  \qquad \sum_{n=1}^{N}A_n^2\le 1 .
\end{equation}
Let $d_{k,n}=\sqrt{(x_k-\tilde x_n)^2+y_k^2+H^2}$ be the UE~$k$--PA~$n$
distance and $d_{0,n}$ the in-guide PA~$n$--feed distance. The effective
uplink channel coefficient of UE~$k$ is
\begin{equation}\label{eq:hk_eff}
  h_k(\tilde{\mathbf x},\boldsymbol\ell^{\mathrm c})=\sum_{n=1}^{N}
  A_n(\boldsymbol\delta)\,\frac{\sqrt{\eta}\,e^{-\alpha_w d_{0,n}}\,e^{-j\phi_{k,n}}}{d_{k,n}},
\end{equation}
where $\eta=c^2/(16\pi^2 f_c^2)$, $\alpha_w$ is the waveguide loss coefficient,
$\phi_{k,n}=\tfrac{2\pi}{\lambda}d_{k,n}+\tfrac{2\pi}{\lambda_g}d_{0,n}$, and
$\lambda_g=\lambda/n_{\mathrm{eff}}$. By Lorentz reciprocity of the passive
waveguide, the downlink extraction cascade of~\cite{Asaad2026DynStatPASS} also
describes the uplink insertion cascade, so~\eqref{eq:hk_eff} follows from the
first-order coupling-cascade approximation. Since the waveguide loss $\alpha_w$ is
far below free-space path loss, placing a radiating point near a user establishes a
strong LoS link. This link lowers the received-power requirement and thereby
improves feasibility.

\subsection{Uplink NOMA with Fixed SIC Order}
All $K$ users transmit simultaneously over the same band; the BS applies SIC. The geometry is actuated once per reconfiguration epoch and held fixed thereafter.
The decoding order is set at the initial configuration, with users relabeled
$k=1,\ldots,K$ so that $|h_{1}|^2\ge\cdots\ge|h_{K}|^2$, and is kept throughout the
optimization; the rates~\eqref{eq:SINR} are achievable under any fixed order, and
optimizing the order jointly with the other variables is left to future work. Under this fixed order, the SINR and offloading rate of UE~$k$ are
\begin{equation}\label{eq:SINR}
  \gamma_k=\frac{p_k|h_k|^2}{\sum_{i=k+1}^{K}p_i|h_i|^2+\sigma^2},
  \qquad R_k^{\mathrm{off}}=B\log_2(1+\gamma_k),
\end{equation}
with $p_k$ the transmit power of UE~$k$ and $\sigma^2$ the noise power. Holding the
order fixed keeps the rates smooth in
$(\tilde{\mathbf x},\boldsymbol\ell^{\mathrm c})$ within the epoch.

\subsection{Computation and Latency Model}
UE~$k$ holds a task of size $L_k$ bits with hard deadline $T_k^{\max}$, and splits
the execution between local processing and remote offloading in parallel. With local
CPU frequency $f_k$, workload $C_k$ cycles/bit, and local rate $R_k^{\mathrm{loc}}=f_k/C_k$,
the resulting completion time is $L_k/(R_k^{\mathrm{loc}}+R_k^{\mathrm{off}})$,
so task completion within the deadline is equivalent to
\begin{equation}\label{eq:latency}
  \frac{f_k}{C_k}+B\log_2(1+\gamma_k)\;\ge\;\frac{L_k}{T_k^{\max}},
  \qquad\forall k .
\end{equation}
Constraint~\eqref{eq:latency} ties feasibility directly to task size and deadline,
rather than to a fixed minimum-rate threshold; the offload
split is implicit in it and not introduced as a separate variable.

\subsection{Power and Energy Efficiency}
UE~$k$ consumes $P_k^{\mathrm{UE}}=p_k+\varepsilon f_k^3+P_k^{\mathrm{cn}}$, with
$\varepsilon$ the switched-capacitance coefficient and $P_k^{\mathrm{cn}}$ a fixed
circuit power. The infrastructure consumes a fixed RF-chain power $P_{\mathrm{cir}}$
and a server power $P_{\mathrm{srv}}(R)$ for aggregate offloaded throughput
$R=\sum_k R_k^{\mathrm{off}}$, where $P_{\mathrm{srv}}$ is non-decreasing and convex;
the specific form used in simulations is given in Section~\ref{sec:results}. The
total system power is
\begin{equation}\label{eq:Ptot}
  P_{\mathrm{tot}}=\sum_{k=1}^{K}\!\big(p_k+\varepsilon f_k^3+P_k^{\mathrm{cn}}\big)
  +P_{\mathrm{cir}}+P_{\mathrm{srv}}\!\Big(\textstyle\sum_{k}R_k^{\mathrm{off}}\Big),
\end{equation}
and the computation energy efficiency, with
$R_k^{\mathrm{tot}}=R_k^{\mathrm{off}}+R_k^{\mathrm{loc}}$, is
$\eta_{\mathrm{EE}}=\sum_k R_k^{\mathrm{tot}}/P_{\mathrm{tot}}$ (bits/Joule).

\subsection{Problem Formulation}
We formulate the joint configuration as an energy-efficiency maximization problem
subject to per-user deadline and system constraints. The problem can be written as
\begin{subequations}\label{eq:P1}
\begin{align}
  \mathcal P_1:\ & \max_{\{p_k,f_k,\tilde x_n,\ell_n^{\mathrm c}\}}\; \eta_{\mathrm{EE}} \label{eq:P1obj}\\
  & \text{s.t.}\; p_k+\varepsilon f_k^3+P_k^{\mathrm{cn}}\le P_k^{\mathrm{th}},\ \forall k, \label{eq:P1b}\\
  & \phantom{\text{s.t.}\;} \tfrac{f_k}{C_k}+B\log_2(1+\gamma_k)\ge \tfrac{L_k}{T_k^{\max}},\ \forall k, \label{eq:P1c}\\
  & \phantom{\text{s.t.}\;} 0\le p_k\le P_k^{\max},\quad 0\le f_k\le f_k^{\max},\ \forall k, \label{eq:P1d}\\
  & \phantom{\text{s.t.}\;} \tilde x_1\ge -\tfrac{D}{2},\ \tilde x_N\le \tfrac{D}{2}, \nonumber\\
  & \phantom{\text{s.t.}\;} \tilde x_{n+1}-\tilde x_n\ge \tfrac{\lambda}{2},\ n\in\{1,\ldots,N-1\}, \label{eq:P1f}\\
  & \phantom{\text{s.t.}\;} \ell_n^{\mathrm c}\in[0,\tfrac{\pi}{2\kappa}],\ |\ell_n^{\mathrm c}-\ell_n^{\mathrm c,\mathrm{prev}}|\le\Delta\ell^{\max},\ \forall n. \label{eq:P1h}
\end{align}
\end{subequations}
Here~\eqref{eq:P1b} is the per-user power budget, \eqref{eq:P1c} the task-completion
deadline, \eqref{eq:P1d} the transmit-power and CPU limits, \eqref{eq:P1f} the PA
placement and minimum-spacing constraints, and~\eqref{eq:P1h} the coupling-length
branch and a per-epoch change limit $|\ell_n^{\mathrm c}-\ell_n^{\mathrm c,\mathrm{prev}}|\le\Delta\ell^{\max}$
that bounds how far each coupling length can be re-actuated between epochs. Problem~$\mathcal P_1$ is a non-convex fractional
program: the objective is fractional, the geometry
$(\tilde{\mathbf x},\boldsymbol\ell^{\mathrm c})$ enters the SINRs non-convexly, and
$P_{\mathrm{srv}}$ makes the denominator non-convex.

Under hard deadlines, feasibility of~\eqref{eq:P1b}--\eqref{eq:P1d} is prior to the
objective: a configuration that meets every deadline within budget must exist
before efficiency can be optimized. Section~\ref{sec:alg} shows that the least-power
feasible point provides a natural initialization for the energy-efficiency
formulation and is especially useful in the deadline-binding regime, which governs
the feasibility behavior in Section~\ref{sec:results}.

\section{Proposed Algorithm}\label{sec:alg}
$\mathcal P_1$ is non-convex, but its variables split into three blocks solved separately: the joint power/CPU block
$(\mathbf p,\mathbf f)$, where $\mathbf p=[p_1,\ldots,p_K]^T$ and
$\mathbf f=[f_1,\ldots,f_K]^T$ stack the per-user transmit powers and CPU
frequencies, the PA-position block $\tilde{\mathbf x}$, and the
coupling-length block $\boldsymbol\ell^{\mathrm c}$. We solve $\mathcal P_1$ by
alternating optimization (AO) over these blocks, initialized at a
least-power feasible point that we characterize first, because it underpins the
design's feasibility behavior.

\subsection{Least-Power Feasible Initialization}\label{sec:init}
Feasibility of the hard deadline~\eqref{eq:P1c} within the power
budget~\eqref{eq:P1b} is a prerequisite to optimizing efficiency. The following
result gives, in closed form, the minimum-power point that meets every deadline
in the pure-offload case and a corresponding feasibility test.

\begin{proposition}[Least-power feasible initialization]\label{prop:init}
With $\mathbf f=\mathbf 0$, constraint~\eqref{eq:P1c} is equivalent to the linear
system $q_k\ge\beta_k^0(\sigma^2+\sum_{i>k}q_i)$ in the received powers
$q_k\triangleq p_k|h_k|^2$, where $\beta_k^0\triangleq 2^{L_k/(BT_k^{\max})}-1$.
Its componentwise-least solution is
\begin{equation}\label{eq:qstar}
  q_k^\star=\sigma^2\beta_k^0\prod_{i=k+1}^{K}(1+\beta_i^0)
          =\sigma^2\beta_k^0\,2^{\sum_{i>k}L_i/(BT_i^{\max})} .
\end{equation}
Setting $p_k^{(0)}=q_k^\star/|h_k|^2$ yields a feasible point if and only if
\begin{equation}\label{eq:init_test}
  p_k^{(0)}\;\le\;\bar P_k\;\triangleq\;\min\{P_k^{\max},\,P_k^{\mathrm{th}}-P_k^{\mathrm{cn}}\},
  \qquad\forall k,
\end{equation}
and this condition is necessary and sufficient for feasibility in the pure-offload
case $\mathbf f=\mathbf 0$, in which every user offloads its entire task with no
local computation.
\end{proposition}

The proof follows by exponentiating~\eqref{eq:P1c} at $f_k=0$ and solving the
resulting triangular system by downward recursion
($q_K^\star=\beta_K^0\sigma^2$, $q_k^\star=\beta_k^0(\sigma^2+\sum_{i>k}q_i^\star)$);
the least-element property follows by induction. Because $\mathbf q^\star$ is the
componentwise-least feasible point, the powers $p_k^{(0)}=q_k^\star/|h_k|^2$
maximize the slack to the effective power budgets $\bar P_k-p_k$. Hence~\eqref{eq:init_test}
certifies feasibility with maximal margin. When a common
nonzero clock $f_k=f^\circ>0$ is admitted, the recursion applies verbatim with
$\beta_k^0$ replaced by $\beta_k(f^\circ)=2^{(L_k/T_k^{\max}-f^\circ/C_k)/B}-1$ and
$\bar P_k$ by $\min\{P_k^{\max},P_k^{\mathrm{th}}-\varepsilon(f^\circ)^3-P_k^{\mathrm{cn}}\}$,
the rescue clocks being capped at the CPU-frequency budget
$f_k^{\mathrm{bud}}\triangleq((P_k^{\mathrm{th}}-P_k^{\mathrm{cn}})/\varepsilon)^{1/3}$
implied by~\eqref{eq:P1b}; if no clock makes~\eqref{eq:init_test} hold, the trial is
declared a
\emph{QoS-feasibility outage}.

\subsection{Joint Power and CPU-Frequency Block}
Fixing the geometry, the power/CPU subproblem maximizes the ratio
$\mathcal N(\mathbf p,\mathbf f)/\mathcal D(\mathbf p,\mathbf f)$ with
 $\mathcal N=B\log_2(1+\sum_k p_k|h_k|^2/\sigma^2)+\sum_k f_k/C_k$ (concave, by the
telescoping identity for the uplink NOMA sum-rate~\cite{zeng2021irs_noma}) and $\mathcal D$
the total power of~\eqref{eq:Ptot}, subject to the deadline~\eqref{eq:P1c} and box
constraints. Power and CPU are solved \emph{jointly} rather than as two separate
blocks, because the deadline~\eqref{eq:P1c} couples $f_k$ and $\mathbf p$
additively within a single per-user constraint, so an independent split risks one
variable invalidating the other's feasibility. The denominator $\mathcal D$ is non-convex because $P_{\mathrm{srv}}$, though convex and
non-decreasing, is composed with the concave aggregate rate. The fractional objective
is handled by a Dinkelbach procedure~\cite{dinkelbach1967} combined with successive
convex approximation~\cite{scutari2017}, alternating between the two variable groups.
With $\mathbf f$ fixed, the deadline~\eqref{eq:P1c} reduces to the linear received-power
constraint $p_k|h_k|^2\ge\beta_k(f_k)\big(\sum_{i>k}p_i|h_i|^2+\sigma^2\big)$; the
server term $P_{\mathrm{srv}}(R(\mathbf p))$ is convexified by linearizing the concave
rate $R(\mathbf p)$ at the current iterate, so the resulting power sub-step is convex.
With $\mathbf p$ fixed, the Dinkelbach objective $\mathcal N-\mu\mathcal D$ separates per user into the concave
$f_k/C_k-\mu\varepsilon f_k^3$, while~\eqref{eq:P1c} and~\eqref{eq:P1b} reduce to
$f_k\in[f_k^{\mathrm{lb}},f_k^{\mathrm{ub}}]$ with
$f_k^{\mathrm{lb}}=C_k\max(0,L_k/T_k^{\max}-R_k^{\mathrm{off}})$ and
$f_k^{\mathrm{ub}}=\min\{f_k^{\max},((P_k^{\mathrm{th}}-P_k^{\mathrm{cn}}-p_k)/\varepsilon)^{1/3}\}$;
hence the exact update is $f_k^\star=\min\{\max\{(3\mu\varepsilon C_k)^{-1/2},f_k^{\mathrm{lb}}\},f_k^{\mathrm{ub}}\}$. The ratio $\mu$ is updated by the standard Dinkelbach rule, updates are
accepted only on non-decrease of $\eta_{\mathrm{EE}}$, and feasibility is preserved
across iterations.

\subsection{PA-Position and Coupling-Length Blocks}
With $(\mathbf p,\mathbf f)$ fixed, the geometry blocks improve the effective
channels $\{|h_k|^2\}$ subject to the placement, spacing, branch, and change-rate
constraints of~\eqref{eq:P1f}--\eqref{eq:P1h}. Both $\tilde{\mathbf x}$ and
$\boldsymbol\ell^{\mathrm c}$ enter $h_k$ non-convexly through~\eqref{eq:hk_eff};
each is optimized by proximal successive convex approximation (SCA)~\cite{scutari2017}.
At iterate $j$, each rate term $R_k^{\mathrm{off}}(\mathbf v)$ (with $\mathbf v$ denoting
$\tilde{\mathbf x}$ or $\boldsymbol\ell^{\mathrm c}$) is replaced by the first-order
surrogate at $\mathbf v^{(j)}$ minus a proximal term $\tfrac{\rho}{2}\|\mathbf v-\mathbf
v^{(j)}\|^2$; since $R_k^{\mathrm{off}}$ is $L$-smooth on the compact domain
($d_{k,n}\ge H>0$), the descent lemma makes this surrogate a global concave minorant
for $\rho\ge L$, so the resulting
strictly concave subproblem yields a monotone, feasibility-preserving update. The
proximal weight is set by backtracking, and updates are accepted only on non-decrease of
$\eta_{\mathrm{EE}}$. The position block is preceded by a grid search to
escape poor stationary points, and the coupling block uses a backtracked step. Each geometry update is accepted only if it preserves feasibility.

\subsection{Overall Algorithm, Convergence, and Complexity}
Algorithm~\ref{alg:AO} summarizes the procedure: the allocation is initialized by
Proposition~\ref{prop:init}; if~\eqref{eq:init_test} fails after the rescue clocks,
an outage is declared; otherwise the three blocks alternate until the
energy-efficiency increment falls below a tolerance.

\begin{algorithm}[t]
\caption{Feasibility-First Joint Resource Allocation}
\label{alg:AO}
\begin{algorithmic}[1]
\REQUIRE Trial $\omega=\{(x_k,y_k),L_k,T_k^{\max},C_k\}_{k=1}^K$; parameters of Table~\ref{tab:sim_params}
\ENSURE $(\mathbf p,\mathbf f,\tilde{\mathbf x},\boldsymbol\ell^{\mathrm c})$, or \textsc{outage}
\STATE \textbf{Initialize geometry:} equispaced $\tilde{\mathbf x}$, $\ell_n^{\mathrm c}=\arccos(1/\sqrt N)/\kappa$; fix SIC order by $|h_k|^2$ (Sec.~\ref{sec:model})
\STATE \textbf{Least-power feasibility (Prop.~\ref{prop:init}):} compute $q_k^\star$ by~\eqref{eq:qstar}, set $p_k^{(0)}=q_k^\star/|h_k|^2$
\IF{$p_k^{(0)}\le\bar P_k\ \forall k$ \eqref{eq:init_test}}
    \STATE feasible; $\mathbf f^{(0)}=\mathbf 0$
\ELSE
    \FOR{$f^\circ\in\{0.25,0.5\}\cdot\min\{f_k^{\max},f_k^{\mathrm{bud}}\}$ (rescue clocks)}
        \STATE recompute $q_k^\star,\bar P_k$ with $\beta_k(f^\circ)$; \textbf{if} \eqref{eq:init_test} holds \textbf{then} feasible; \textbf{break}
    \ENDFOR
    \STATE \textbf{if} no clock feasible \textbf{then} \textbf{return} \textsc{outage}
\ENDIF
\REPEAT
        \STATE \textbf{Step 1 (power/CPU):} update $(\mathbf p,\mathbf f)$ by Dinkelbach with SCA~\cite{dinkelbach1967,scutari2017}
    \STATE \textbf{Step 2 (positions):} grid search then SCA~\cite{scutari2017} on $\tilde{\mathbf x}$; accept only if $\eta_{\mathrm{EE}}$ non-decreasing
    \STATE \textbf{Step 3 (coupling):} backtracked SCA on $\boldsymbol\ell^{\mathrm c}$; accept only if $\eta_{\mathrm{EE}}$ non-decreasing
\UNTIL{$\eta_{\mathrm{EE}}$ increment $<\epsilon_{\mathrm{tol}}$}
\RETURN $(\mathbf p,\mathbf f,\tilde{\mathbf x},\boldsymbol\ell^{\mathrm c})$
\end{algorithmic}
\end{algorithm}

\begin{theorem}\label{thm:conv}
The sequence $\{\eta_{\mathrm{EE}}^{(i)}\}$ generated by
Algorithm~\ref{alg:AO} is monotonically non-decreasing and convergent.
\end{theorem}

Monotonicity holds because each block preserves or increases $\eta_{\mathrm{EE}}$:
each convex-surrogate step is accepted only on non-decrease, so no block lowers the
objective; the sequence
is bounded above (finite power budgets), hence convergent. With $I_{\mathrm{AO}}$, $I_{\mathrm D}$, $I_{\mathrm S}$ the outer-AO, Dinkelbach, and
SCA iteration counts, $|\mathcal G|$ the grid size, and each surrogate solved by an
interior-point method ($O(n^{3.5})$ in the $n$ block variables), the per-run cost is
polynomial in $K$ and $N$:
\begin{equation}\label{eq:complexity}
  \mathcal O\!\big(I_{\mathrm{AO}}\big[I_{\mathrm D}K^{3.5}+|\mathcal G|KN^2
  +I_{\mathrm S}(KN^2+N^{3.5})\big]\big).
\end{equation}

\section{Simulation Results and Discussion}\label{sec:results}

\subsection{Setup and QoS-Feasibility Outage}\label{sec:setup}
We evaluate the design under the parameters of Table~\ref{tab:sim_params}, with $N=8$ PAs unless stated otherwise. Each Monte Carlo trial places $K$ users independently and
uniformly in the service region
($x_k\in[-10,10]$~m, $y_k\in[1,10]$~m) with task sizes
$L_k\sim\mathcal U(2,8)$~Mbit and a common deadline $T^{\max}=50$~ms; rates are
reported in Mbit/s with $\tilde B\triangleq B/10^6=100$. The server power follows the
cubic model $P_{\mathrm{srv}}(R)=\varepsilon_{\mathrm{srv}}(C_{\mathrm{srv}}R)^3$. We compare four schemes:
the \emph{proposed} design
(Algorithm~\ref{alg:AO}); \emph{single-PA}, a lone fully-radiating antenna optimized
by the same engine; \emph{OMA-MEC}, interference-free TDMA; and \emph{equal-power},
without power control, each evaluated by the same outage criterion.

\begin{definition}[QoS-feasibility outage]\label{def:outage}
For feasible set $\mathcal F(\omega)$ of $\mathcal P_1$ under trial $\omega$, the
QoS-feasibility outage probability is
\begin{equation}\label{eq:outage}
  P_{\mathrm{out}}\triangleq\Pr\big\{\mathcal F(\omega)=\varnothing\big\}
  =\Pr\big\{\nexists\,(\mathbf p,\mathbf f)\ \text{satisfying~\eqref{eq:P1b}--\eqref{eq:P1d}}\big\},
\end{equation}
estimated by $\hat P_{\mathrm{out}}$, the fraction of trials for which the
test~\eqref{eq:init_test}, at the initialized geometry with $\mathbf f=\mathbf 0$ and
the rescue clocks, fails. A passing trial is certified feasible, so
$\hat P_{\mathrm{out}}\ge P_{\mathrm{out}}$, with equality when restricted to pure
offloading.
\end{definition}
$P_{\mathrm{out}}$ is the complement of the task-completion probability standard in
deadline-constrained MEC and is distinct from SINR outage; all reported outage values
are therefore conservative.

\begin{table}[t]
\centering
\caption{Simulation Parameters}
\label{tab:sim_params}
\begin{tabular}{ll}
\toprule
Parameter & Value \\
\midrule
Carrier $f_c$, bandwidth $B$ & 28 GHz, 100 MHz \\
Waveguide length $D$, height $H$ & 20 m, 3 m \\
PA count $N$ (swept) & $\{1,2,4,8,16\}$ \\
UE count $K$ (swept) &$\{2,\ldots,12\}$ \\
UE region & $x_k\in[-10,10]$ m, $y_k\in[1,10]$ m \\
Waveguide loss $\alpha_w$, coupling $\kappa$ & 0.08 Np/m, 15.7 rad/m \\
Task size $L_k$, deadline $T_k^{\max}$ & $\mathcal U(2,8)$ Mbit, 50 ms \\
CPU cycles/bit $C_k$ & $\mathcal U\{500,\ldots,1000\}$ \\
$\varepsilon$, $P_k^{\mathrm{cn}}$, $P_{\mathrm{cir}}$ & $10^{-27}$, 100 mW, 1 W \\
$P_k^{\max}$, $P_k^{\mathrm{th}}$, $f_k^{\max}$ & 200 mW (nominal), 500 mW, 1 GHz \\
$\varepsilon_{\mathrm{srv}}$, $C_{\mathrm{srv}}$ & $10^{-31}$, 100 cyc/bit \\
Noise power $\sigma^2$ & $-90$ dBm over $B$ (incl.\ 4 dB NF) \\
\bottomrule
\end{tabular}
\end{table}

\subsection{Feasibility versus Number of Users}\label{sec:outageK}
\begin{figure}[t]
\centering
\includegraphics[width=\columnwidth]{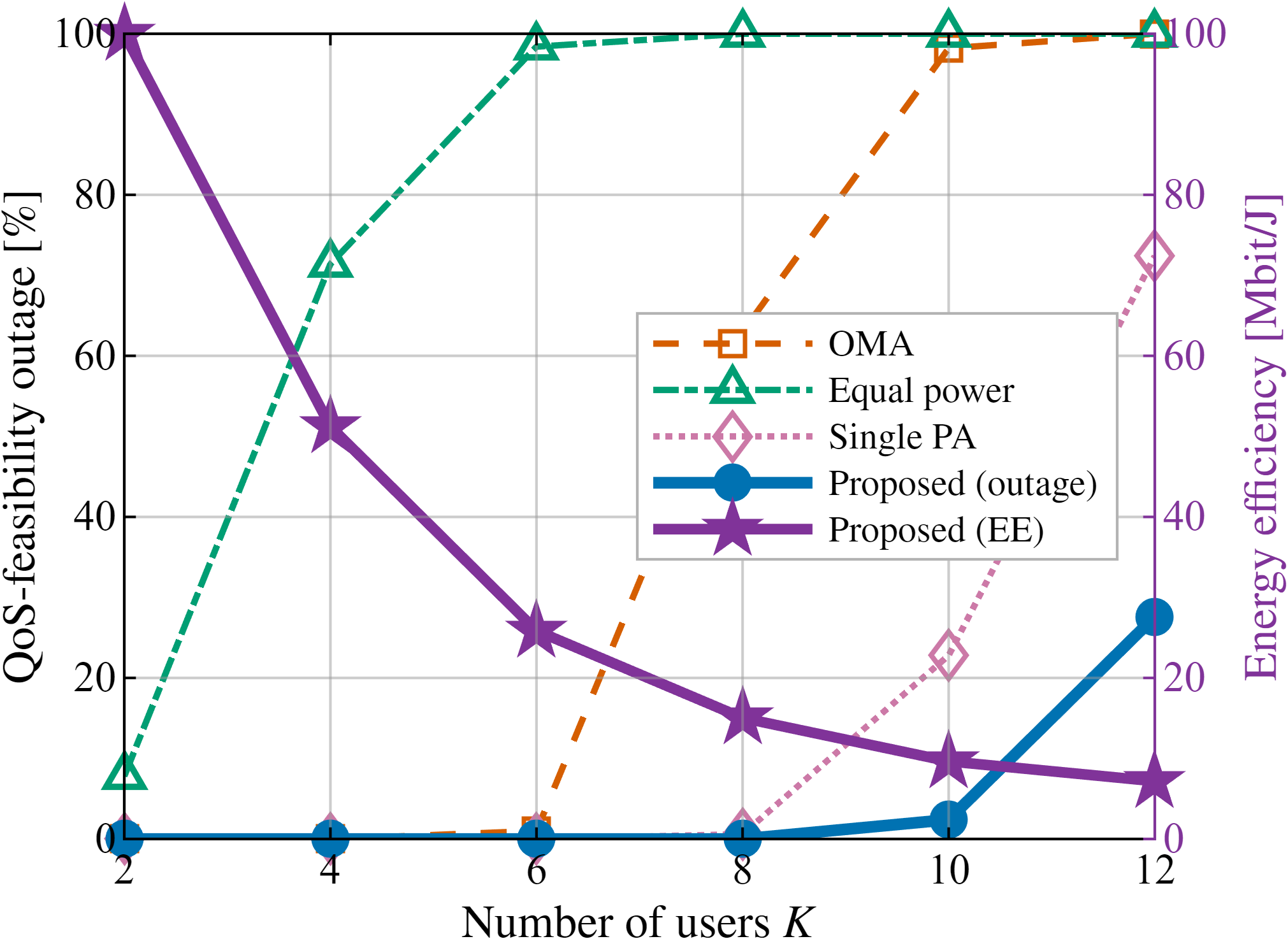}
\caption{QoS-feasibility outage (left axis) and energy efficiency of the proposed
design (right axis) versus number of users $K$ (average over $500$ trials per point).}
\label{fig:outage}
\end{figure}

Fig.~\ref{fig:outage} reports QoS-feasibility outage versus $K$ (average over $500$ trials per point), the central result of this paper. We sweep $K\ge2$, since NOMA requires at least two superposed users. The proposed design admits all users with
zero observed outage through $K=8$ and degrades gracefully thereafter ($2.4\%$ at $K=10$,
$27.6\%$ at $K=12$), the onset occurring only when the aggregate received-power
requirement $q_k^\star\approx2^{\sum_{i\ge k}R_i^{\mathrm{req}}/\tilde B}$
of~\eqref{eq:qstar}, with $R_i^{\mathrm{req}}\triangleq L_i/T_i^{\max}$, exceeds the budget for the weakest user. The baselines collapse
earlier, each by a distinct and model-predicted mechanism.

\emph{Equal power} fails between $K=4$ and $K=6$: without power control the
first-decoded rate is $\tilde B\log_2(K/(K\!-\!1))$, which falls from $41.5$~Mbit/s
at $K=4$ to $26.3$~Mbit/s at $K=6$, crossing the $40$~Mbit/s deadline floor and
producing $71.4\%$ then $98.4\%$ outage. \emph{OMA-MEC} fails between $K=8$ and $K=10$: equal time-sharing compresses
each user's slot to $1/K$, inflating the per-user requirement to
$2^{KR_k^{\mathrm{req}}/\tilde B}-1$, which breaches the $200$~mW budget ($63\%$ outage at $K=8$).
The \emph{single fully-radiating PA} is the most robust baseline, nearly matching the
proposed design through $K=8$ ($0.6\%$), but its single path cannot serve the far,
high-demand users as $K$ grows, yielding $22.8\%$ and $72.4\%$ outage at $K=10$ and
$K=12$. The right axis reports the energy efficiency, the quantity
$\mathcal P_1$ maximizes: highest at light load, it decreases as $K$
grows since more users share the fixed infrastructure power. The design thus
optimizes its stated objective while sustaining feasibility, the governing metric of
this work.

\subsection{Feasibility under Power-Budget Stress}\label{sec:budget}
\begin{figure}[t]
\centering
\includegraphics[width=\columnwidth]{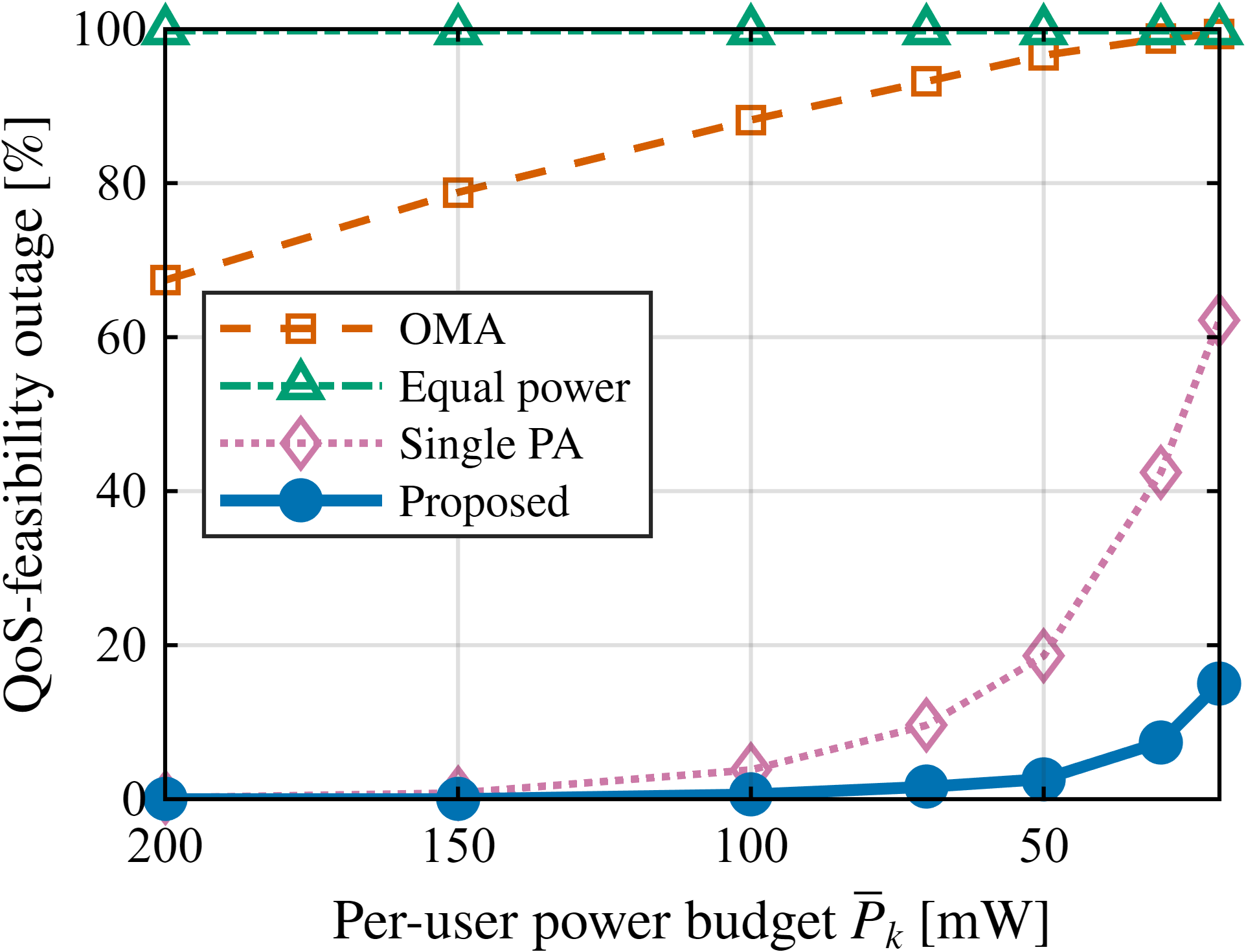}
\caption{QoS-feasibility outage versus per-user power budget $\bar P_k$ at $K=8$
(average over $500$ trials per point).}
\label{fig:budget}
\end{figure}

Fig.~\ref{fig:budget} fixes $K=8$ and sweeps the per-user power budget
$\bar P_k$, tracing the feasibility margin of Proposition~\ref{prop:init}
directly: as $\bar P_k$ shrinks, the least-power point
$p_k^{(0)}=q_k^\star/|h_k|^2$ breaches the budget for more trials, so outage rises.
The proposed design degrades gracefully, its outage never exceeding $15\%$ even
as the budget is cut by an order of magnitude (from $200$ to $20$~mW), because the
multi-antenna configuration distributes the received-power requirement. The
single-antenna design matches it at generous budgets but collapses under stress
($62.2\%$ at $20$~mW), isolating the feasibility contribution of the
pinching-antenna array; OMA and equal power already fail on most trials
($67.4\%$ and $99.8\%$) at the nominal budget. 

\subsection{Feasibility versus Array Size}\label{sec:arraysize}
\begin{figure}[t]
\centering
\includegraphics[width=\columnwidth]{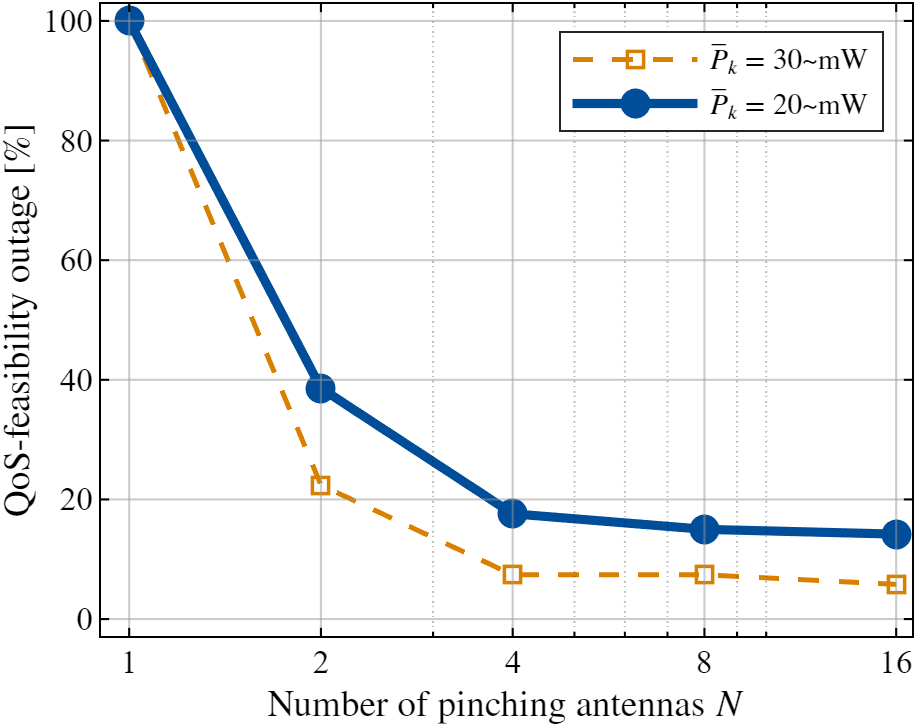}
\caption{QoS-feasibility outage versus number of pinching antennas $N$ at $K=8$
(average over $500$ trials per point).}
\label{fig:arraysize}
\end{figure}

The preceding results fix the antenna array and vary the load; we now fix the load
and vary the array. Fig.~\ref{fig:arraysize} sweeps the number of pinching antennas
$N$ at $K=8$ under constrained budgets. A single element ($N=1$) cannot
serve the eight users at either budget (outage is $100\%$), whereas adding antennas
drives feasibility up sharply: at $\bar P_k=20$~mW outage falls to $17.6\%$ at $N=4$
and $15\%$ at $N=8$, and at $\bar P_k=30$~mW it falls to $7.4\%$ at $N=8$ and
$5.8\%$ at $N=16$. The mechanism is that of Proposition~\ref{prop:init}: each
additional radiating point lowers the per-user received-power requirement
$q_k^\star/|h_k|^2$, enlarging the margin to the budget so that the least-power test
holds for more trials. This isolates the feasibility contribution of the
pinching-antenna array itself, distinct from the power allocation. The gain saturates
beyond $N=8$ at the tighter budget, where the residual outage is budget- rather than
array-limited. The array size improves feasibility, while fine positioning and coupling refine the
energy-efficiency operating point within the feasible set; feasibility itself is
governed by the least-power test of Proposition~\ref{prop:init} and the number of
radiating points.

\subsection{Convergence and Complexity}\label{sec:convresult}
At $K=8$, Algorithm~\ref{alg:AO} terminates in a median and maximum of two outer
iterations with no monotonicity violation, agreeing with Theorem~\ref{thm:conv};
the efficiency increment was contained almost entirely in the first joint power/CPU
update, consistent with the near-optimality of the least-power initializer under the
binding deadlines~\eqref{eq:P1c}. With these small iteration counts, the polynomial
per-run cost~\eqref{eq:complexity} solves a full trial in a fraction of the per-epoch reconfiguration timescale.
\section{Conclusion}\label{sec:concl}
We studied deadline-constrained computation offloading in an uplink NOMA-MEC system
served by a pinching-antenna system, framing the design around \emph{feasibility}:
sustaining every user's hard task-completion deadline within per-user power budgets
as the number of users grows. We formulated the joint optimization of transmit
powers, CPU frequencies, and antenna configuration with energy efficiency as the
objective and the deadlines as constraints, and derived a closed-form least-power
feasible initialization that certifies feasibility with the largest attainable margin to the
power budgets and yields a QoS-feasibility test. A three-block alternating-optimization algorithm with proven monotone convergence and polynomial complexity completes the solution. Numerical results showed that the proposed design
admits all users with zero QoS-feasibility outage through $K=8$ and degrades
gracefully thereafter, whereas equal-power, OMA, and single-antenna baselines fail
earlier. Stress tests
further showed that the design sustains feasibility under order-of-magnitude budget
reductions and that the pinching-antenna array itself enables
constrained-budget operation. Unlike recent
DRL~\cite{ma2026drl_pass_mec,hu2026pass_mec} and delay-minimizing~\cite{ai2026delay_pass_mec}
PASS-MEC designs, our framework offers a $K$-user feasibility certificate.
\section*{Acknowledgment}
The work of Jiguang He is partially supported by the Funding Program for High-Level Overseas-Educated Talents (Grant No. RCXMA26001) and the Guangdong Provincial Project (Grant No. 2024QN11X433).

\bibliographystyle{IEEEtran}
\bibliography{references}

@article{fukuda2022pinching,
  title     = {Pinching Antenna---Using a Dielectric Waveguide as an Antenna},
  author    = {Fukuda, Atsushi and Yamamoto, Hiroto and Okazaki, Hiroshi and Suzuki, Yasunori and Kawai, Kunihiro},
  journal   = {NTT DOCOMO Technical Journal},
  volume    = {23},
  number    = {3},
  pages     = {5--12},
  year      = {2022},
  month     = jan,
  publisher = {NTT DOCOMO, Inc.},
}

@ARTICLE{liu2026pass_tutorial,
  author={Liu, Yuanwei and Jiang, Hao and Xu, Xiaoxia and Wang, Zhaolin and Guo, Jia and Ouyang, Chongjun and Mu, Xidong and Ding, Zhiguo and Nallanathan, Arumugam and Karagiannidis, George K. and Schober, Robert},
  journal={IEEE Transactions on Communications}, 
  title = {Pinching-antenna systems ({PASS}): A tutorial}, 
  year={2026},
  volume={74},
  number={},
  pages={4881-4918},
  doi={10.1109/TCOMM.2026.3658289}}

@ARTICLE{tyrovolas2025performance,
  author={Tyrovolas, Dimitrios and Tegos, Sotiris A. and Diamantoulakis, Panagiotis D. and Ioannidis, Sotiris and Liaskos, Christos K. and Karagiannidis, George K.},
  journal={IEEE Transactions on Cognitive Communications and Networking}, 
  title={Performance Analysis of Pinching-Antenna Systems}, 
  year={2026},
  volume={12},
  number={},
  pages={590-601},
  doi={10.1109/TCCN.2025.3564470}}

@ARTICLE{zeng2025ee_noma_pass,
  author={Zeng, Ming and Li, Xingwang and Wang, Ji and Huang, Gaojian and Dobre, Octavia A. and Ding, Zhiguo},
  journal={IEEE Wireless Communications Letters}, 
  title={Energy-Efficient Resource Allocation for {NOMA}-Assisted Uplink Pinching-Antenna Systems}, 
  year={2025},
  volume={14},
  number={11},
  pages={3695-3699},
  doi={10.1109/LWC.2025.3600899}}

@ARTICLE{tegos2025uplink_pass,
  author={Tegos, Sotiris A. and Diamantoulakis, Panagiotis D. and Ding, Zhiguo and Karagiannidis, George K.},
  journal={IEEE Wireless Communications Letters}, 
  title={Minimum Data Rate Maximization for Uplink Pinching-Antenna Systems}, 
  year={2025},
  volume={14},
  number={5},
  pages={1516-1520},
  doi={10.1109/LWC.2025.3547956}}

@INPROCEEDINGS{wang2022irs_mec_noma,
  author={Wang, Qun and Zhou, Fuhui and Hu, Han and Hu, Rose Qingyang},
  booktitle={2021 13th International Conference on Wireless Communications and Signal Processing (WCSP)}, 
  title = {Energy-efficient design for {IRS}-assisted {MEC} networks with {NOMA}}, 
  year={2021},
  volume={},
  number={},
  pages={1-6},
  doi={10.1109/WCSP52459.2021.9613291}}

@ARTICLE{zuo2024fluid_mec,
  author={Zuo, Yiping and Guo, Jiajia and Sheng, Biyun and Dai, Chen and Xiao, Fu and Jin, Shi},
  journal={IEEE Communications Letters}, 
  title={Fluid Antenna for Mobile Edge Computing}, 
  year={2024},
  volume={28},
  number={7},
  pages={1728-1732},
  doi={10.1109/LCOMM.2024.3399407}}

@ARTICLE{zhu2024movable,
  author={Zhu, Lipeng and Ma, Wenyan and Zhang, Rui},
  journal={IEEE Communications Magazine}, 
  title={Movable Antennas for Wireless Communication: Opportunities and Challenges}, 
  year={2024},
  volume={62},
  number={6},
  pages={114-120},
  doi={10.1109/MCOM.001.2300212}}

@INPROCEEDINGS{Asaad2026DynStatPASS,
  author={Asaad, Saba and Ouyang, Chongjun and Bereyhi, Ali and Ding, Zhiguo},
  booktitle={ICC 2026 - IEEE International Conference on Communications}, 
  title={Dynamic and Static Energy Efficient Design of Pinching Antenna Systems}, 
  year={2026},
  volume={},
  number={},
  pages={1-6},
  doi={10.1109/ICC59461.2026.11586995}}

@ARTICLE{ma2026drl_pass_mec,
  author={Ma, Yujie and Huang, Wenqi and Ding, Zhiguo and So, Daniel K. C. and Bhatia, Vimal},
  journal={IEEE Wireless Communications Letters}, 
  title = {Joint optimization in pinching-antenna-assisted {NOMA}-{MEC} systems via {DRL}}, 
  year={2026},
  volume={15},
  number={},
  pages={2493-2497},
  doi={10.1109/LWC.2026.3680977}}

@article{dinkelbach1967,
  title={On nonlinear fractional programming},
  author={Dinkelbach, Werner},
  journal = {Management Science},
  volume={13},
  number={7},
  pages={492--498},
  year={1967},
  publisher={INFORMS}
}

@ARTICLE{scutari2017,
  author={Scutari, Gesualdo and Facchinei, Francisco and Lampariello, Lorenzo},
  journal={IEEE Transactions on Signal Processing}, 
  title={Parallel and Distributed Methods for Constrained Nonconvex Optimization—Part I: Theory}, 
  year={2017},
  volume={65},
  number={8},
  pages={1929-1944},
  doi={10.1109/TSP.2016.2637317}}

@ARTICLE{ai2026delay_pass_mec,
  author={Ai, Yuan and Mu, Xidong and Si, Pengbo and Liu, Yuanwei},
  journal={IEEE Communications Letters}, 
  title = {Delay minimization in pinching-antenna-enabled {NOMA}-{MEC} networks}, 
  year={2026},
  volume={30},
  number={},
  pages={962-966},
  doi={10.1109/LCOMM.2026.3655158}}

@ARTICLE{zeng2021irs_noma,
  author={Zeng, Ming and Li, Xingwang and Li, Gen and Hao, Wanming and Dobre, Octavia A.},
  journal={IEEE Communications Letters}, 
  title = {Sum rate maximization for {IRS}-assisted uplink {NOMA}}, 
  year={2021},
  volume={25},
  number={1},
  pages={234-238},
  doi={10.1109/LCOMM.2020.3025978}}

@INPROCEEDINGS{hu2026pass_mec,
  author={Hu, Zhaoming and Zhong, Ruikang and Mu, Xidong and Li, Dengao and Liu, Yuanwei},
    booktitle={Proc. IEEE Int. Conf. Commun. (ICC)},
  title={{PASS}-Enhanced {MEC}: Joint Optimization of Task Offloading and Uplink {PASS} Beamforming},
  year={2026},
  volume={},
  number={},
  pages={1-6},
  doi={10.1109/ICC59461.2026.11587765}}
\end{document}